# Experimental consequences at high temperatures of quantum critical points

D. C. Freitas[1], P. Rodière[1], M. Núñez[2], J. Marcus[1], F. Gay[1], M. A. Continentino[3] and M. Núñez-Regueiro[1*]

1. *Institut Néel, Université Grenoble Alpes & CNRS, 25 Av. des Martyrs, 38042 Grenoble, France*

2. *Consejo Nacional de Investigaciones Científicas y Técnicas, Buenos Aires, Argentina*

*Instituto de Ciencias Básicas, Universidad Nacional de Cuyo, Mendoza, Argentina*

*Departamento Materiales Nucleares, Centro Atómico Bariloche, Comisión Nacional de Energía Atómica, 8400 Bariloche, Argentina*

3. *Centro Brasileiro de Pesquisas Físicas, Rua Dr. Xavier Sigaud, 150, Rio de Janeiro - RJ - 22290-180, Brazil*

\* to whom correspondence should be addressed.

E-mail: manolo.nunez-regueiro@neel.grenoble.fr

**ABSTRACT**

We study the $Cr_{1-x}Re_x$ phase diagram finding that its phase transition temperature towards an antiferromagnetic order $T_N$ follows a quantum $[(x_c - x)/x_c]^\psi$ law, with $\psi = 1/2$, from the quantum critical point (QCP) at $x_c = 0.25$ up to $T_N \approx 600K$. We compare this system to others in order to understand why this elemental material is affected by the QCP up to such unusually high temperatures. We determine a general criterion for the crossover, as function of an external parameter such as concentration, from the region controlled solely by thermal fluctuations to that where quantum effects become observable. The properties of materials with low coherence lengths will thus be altered far away from the QCP.

Theoretical studies by Hertz [1] and Millis [2] (HM) have directed research towards quantum phase transitions (QCP) [3,4] induced by an external parameter $\delta$, such as doping, pressure or magnetic field [5,6]. In thermal phase transitions the critical temperature results from the competition between an interaction that drives the system to an ordered state and the entropy, which increases in the presence of thermal disorder. As QCP occur at zero temperature there is no entropy, and it is now the competition among the different terms of the Hamiltonian describing the material which gives rise to the QCP. In an itinerant magnetic system this competition involves the Coulomb repulsion $U$ between electrons and the electronic kinetic energy as expressed by the bandwidth $W$. An increase of the ratio $U/W$ due to pressure or doping can thus set up a magnetic ground state at a QCP. In three dimensions this ground state is robust to thermal fluctuations and exists up to a critical temperature, which has a power law dependence with the distance to the QCP. QCP's have been studied in heavy fermions [7] whose characteristic energies rarely exceed 30K and the present theory has successfully interpreted the main experimental results. On the other hand, although calculations on ideal systems have predicted observation of quantum critical behavior over an extended temperature range [8], experimental evidence is needed to support these theoretical predictions. Furthermore, few works have been addressed the passage from classical to quantum dominated regions of the phase diagram of elemental materials.

Chromium presents a spin density wave (SDW) due to the nesting of the hole and electron pockets of its Fermi surface that develops at a Néel temperature $T_N = 312K$. $Cr$ metal displays two clearly defined regimes on its way to the QCP both as a function of pressure $P$ or vanadium doping $x$ ($Cr_{1-x}V_x$). A classical region at low $P$ or $x$, where $T_N$ decreases following an exponential with pressure BCS tuned variation [9] consequence of the dependence of $T_N$ with the coupling parameter $\lambda$, i.e. $T_N \sim e^{-1/\lambda}$. A quantum region appears below a crossover value [10,11] of $T_N$ where quantum effects take control through a power law behavior with an exponent $\psi = \frac{1}{2}$, the same for $P$ and $x$.

Partial substitution of $Cr$ by $Re$ increases the electron count of the alloy. It first enhances the nesting and thus augments $T_N$, then decreases it and at $x \sim 0.15$ superconductivity appears [12], coexisting with antiferromagnetism. As metallurgical drawbacks cast doubts on these reported results, we have determined a detailed phase diagram with more homogenously sputtered samples (Supplemental Material) of the $Cr_{1-x}Re_x$ system.

The sputtered series of samples was grown using a magnetron sputtering apparatus with characteristics similar to those of Ref. 13 , with a base pressure below $5*10^{-8}$ Torr. The $Cr$ sputtering target was covered with appropriate angular portions of $Re$ 0.2mm foil to the desired concentration. The gun was powered with 150 W dc, and $Cr_{1-x}Re_x$ films were deposited at rates of 0.5–1.0 Å/ s onto substrates of amorphous SiN$x$ / Si held at room temperature. The Ar sputtering gas pressure was approximately 0.75 mTorr. The $Re$ concentration was verified by scanning electron microscope (details in Supplemental Material).

In bulk samples, it is difficult to avoid concentration gradients because of the high liquidus temperature of the alloys (above 2000°C) together to the high vapor pressure of Cr at elevated temperatures [14]. This metallurgical drawback is amplified at higher Re concentrations (approximately $0.2 < x_{Re} < 03$), allowing phase separations, which can explain the coexistence of low Re content regions (having a $T_N \approx 157K$) with high Re regions (superconducting with $T_{sc} \approx 2.7K$) in the bulk samples. The sputtering technique circumvents these metallurgical problems by creating a homogeneous plasma with the Cr/Re ratio of the target, which is then deposited at room temperature into a film with the corresponding homogeneity. The very quick annealing does not affect this initial homogeneity, as confirmed by our scanning electron microscope photos (Fig. 1 Supp.).

The electrical resistivity was measured using a four lead direct current method. Depending on the temperature range the contacts were tungsten fingers or platinum leads with silver epoxy. Annealing and high temperature electrical resistivity measurements were done simultaneously in a home-made optical furnace with four lead tungsten contacts. The samples were heated up to 800°C in a ten minutes lapse and measured down to room temperature at the same speed. Low temperature measurements were done on $^3$He and $^4$He cryostats.

On Fig. 1 we show the electrical resistances of the samples as a function of temperature. The characteristic anomaly due to the SDW is clearly seen on the resisitivity curves. We determine the Néel temperature $T_N$ by the peak of the temperature derivative of the resistance as usual (Supplemental material) and show on Fig. 2 the obtained values as a function of *Re* substitution together with other reported values [12,15]. Far from the QCP, our values agree with precedent reports. Beyond $x \approx 0.2$, we do not confirm a constant value for $T_N$. Contrary to previous reports [12,15] (triangles on Fig. 2), we do not observe coexistence between magnetism and superconductivity down to 400mK. Above $x \approx 0.25$ the superconducting transition temperature $T_{sc}$ increases monotonically. The difference is certainly due to the different sample homogeneity (see Supplemental Material). It is clear from our results that superconductivity is quenched in the region where Cr is antiferromagnetic. Consequently, the observed superconductivity is presumably conventional electron-phonon superconductivity due to bcc *Re* similar to that observed in W/Re and Mo/Re alloys [12]. The superconducting dome frequent at QCPs [7] is absent.

We now search in the phase diagram of Fig. 2 for the classical and the quantum regimes obtained for *Cr* under pressure or *V* doping. The non-monotonic behavior with doping and the incommensurate to commensurate SDW transition [15] at $x \approx 0.02$ make it difficult to define the classical exponential with doping regime at low Re doping. While the quantum regime is clearly visible for doping concentrations above $x \approx 0.1$ as $T_N(x)$ follows a $[(x_c - x)/x_c]^\psi$ law with $\psi = 1/2$. Our measurements indicate that the quantum effects expected to be important only at low temperatures control the variation of $T_N$ from ~600K down to the QCP. We understand this as follows. The finite temperature transition at $T_N$ is a classical critical phenomenon that arises in renormalization group (RNG) approach. RNG shows that a finite temperature fixed point always governs the critical behavior along the critical line, i.e. temperature is a relevant field at the QCP. For chromium and its alloys the same parameters govern the phase diagram in the whole pressure range. However, while the BCS exponential behavior results from a competition with entropic effects, the power law behavior is a consequence of purely quantum effects. Then, as long as $T_N$ depends on the distance to the QCP its dependence is reflecting the competition between the quantum effects that gave rise to the QCP. In this sense the power law behavior of $T_N$ with the distance to the QCP is a direct evidence of quantum effects up to ~600K. Unfortunately, the study of criticality for other physical quantities, Hall effect and increase of the resistance at $T_N$, was impossible due to two-band effects and domains, respectively (see Supplemental material).

The free energy density including quantum effects within mean field of a magnetic system that at $T = 0K$ will develop a QPT at an external parameter $\delta_c$ is

$$F = \tfrac{1}{2}\alpha(\delta - \delta_c)M^2 + \tfrac{1}{4}uM^4 + \tfrac{1}{2}\xi_{coh}^2(\nabla M)^2 \qquad (1)$$

where $M$ is the order parameter, $\xi_{coh}$ the coherence length, $\alpha$ and $u$ constants. Temperature introduces essentially two corrections to the critical external parameter

$$\delta_c^*(T) = \delta_c - bT^2 - uT^{1/\psi}$$

The $T^2$ term arises from an analytic expansion of the free energy in powers of the temperature (a short calculation is described in Supplemental material). It is always present[16,17] regardless the character, fermionic[16] or bosonic[17], of the critical modes close to the QPT. It is also independent of the universality class of the transition, and does not convey any information on the nature of the QPT. The energy/time HM treatment yields the non-analytic term $T^{1/\psi}$ that is due[4] to the quartic interaction ($uM^4$) which is dangerously irrelevant for systems with effective dimensions $d + z > d_{eff}$, where $d_{eff} = 4$ for magnetic transitions. Here $z$ is the dynamic exponent that takes the value $z = 2$ for a nearly antiferromagnetic metal, and $d$ is the dimensionality. The shift exponent that depends on the universality class of the quantum phase transition can be expressed in terms of these later quantities as $\psi = (d + z - 2)/z$. Then, for an itinerant system in 3d near an antiferromagnetic QCP we expect $\psi = 3/2$ so that at sufficiently low temperatures it always determines the shape of the critical line. However, as $\frac{1}{\psi} < 2$, this may occur only at very low temperatures, especially if the quartic interaction $u$ is very small. For chromium alloys (and the cases that will be discussed below) the analytic $T^2$ term seems to be always the dominant one. Furthermore, the range in temperature of the validity of this law gives strong evidence that its origin is as discussed above and not from unknown fluctuations with dynamic exponent z=1, which in 3d leads to $\psi = 1/2$.

But what is that makes that the transition towards the SDW in $Cr_{1-x}Re_x$ is already under the influence of the QCP at such high temperatures? The answer is inherent in the third term of eq. 1 implying that systems with large coherence lengths will be harder to deform and presumably less sensitive to quantum effects, hypothesis that is supported by the following facts.

The range of observation of the square root power law varies from $T_N \sim 60K$ for pure chromium to $T_N \sim 600K$ for $Cr_{1-x}Re_x$ alloys, suggesting that an increase of disorder causes the observation of quantum effects at higher temperatures. There are at least two other compounds that show a behavior with pressure similar to that of $Cr$ metal. The one dimensional transition metal trichalcogenide $NbSe_3$, a well-studied charge density wave (CDW) compound [18,19,20] and $\beta - Na_{0.33}V_2O_5$, a material with a charge order (CO) [21]. As shown on the left panel of Fig. 3 at low pressures these compounds follow the classical exponential with pressure dependence. At a certain, different for each material, value of the transition temperature, $T_{CDW} \sim 90K$ for $NbSe_3$, $T_{CO} \sim 30K$ for $\beta - Na_{0.33}V_2O_5$, and $T_N \sim 60K$ for $Cr$, the quantum regime sets in.

The above described conventional behavior contrasts with the one shown by the four materials on the right panel of Fig. 3. Layered $1T - TiSe_2$ develops a charge density wave thought to be the result of a transition towards an excitonic insulator at 200K at ambient pressure [22]. One dimensional $o - TaS_3$ develops a CDW at 215K at ambient pressure [23] and the layered high temperature superconducting pnictide $BaFe_2As_2$ an antiferromagnetic ordering at 135K at ambient pressure [24]. All three follow the $\left[(P_c - P)/P_c\right]^{0.5}$ law in the entire pressure phase diagram, as does $Cr_{1-x}Re_x$ as a function of concentration.

The simplest quantity that sorts the materials into the corresponding class, is the measured zero temperature coherence length at ambient pressure or zero concentration $\xi_{meas}$, either experimentally obtained or estimated from domain size (Table I). In $NbSe_3$, both from X-rays measurements [25] or non-linear conductance [26] (that yield the depinning domains), at low temperatures the correlation length has a very large size, ~10000Å. While a lower limit [27] for domains in $\beta - Na_{0.33}V_2O_5$ is 1000Å. Whereas the correlation length [28] as measured by X-rays diffraction at 5K in $Cr$ is well over 2000Å. In contrast to $NbSe_3$ for $o - TaS_3$ the correlation length can be roughly estimated from X-rays measurements [29] at about 500Å. Different chirality [30] renders domains of ~100Å in $1T - TiSe_2$, while the measured magnetic correlation length [31] of $BaFe_2As_2$ at 3K is ~350Å. Finally, a crude lower estimate for $Cr_{1-x}Re_x$ in our measurement range can be given by the distance between Re atoms, ~15Å, as we are in the dirty limit the domain coherence length will certainly be limited by the impurities (Re atoms) in this case.

The compounds on the left panel of Fig. 3 have all $\xi_{meas} > 500 \text{Å}$ while the materials on the right panel have all $\xi_{meas} < 500 \text{Å}$. Consequently, quantum effects dominate the entire phase diagram of materials with small correlation lengths, either intrinsic or impurity controlled. This argument can be quantified further through a simple calculation, showed graphically on Fig. 4.

There are two regimes: classical and quantum, and we want to determine when the system crossovers from one to the other. For the classical regime the transition temperature is, according to BCS, $T_c^{class} = \hbar\omega_0 e^{-1/\lambda}$, where $\omega_0$ is a characteristic energy and $\lambda$ the coupling parameter. The BCS coherence length is $\xi_{coh} = v_F/\Delta$, where $v_F$ is the velocity at the Fermi level and $\Delta$ the BCS gap, proportional to $T_c^{class}$. For the quantum regime the transition temperature is given by $T_c^{quan} = T_{c_0}^{quan}\left[(\delta_c - \delta)/\delta_c\right]^{1/2}$, while the quantum correlation length is $\xi_q = \xi_0[(\delta_c - \delta)/\delta_c]^{-1/2}$. The system changes regime when $T_c^{quan} = T_c^{class}$, thus the crossover occurs when the ratio $\xi_{coh}/\xi_q = \kappa$, a constant particular to the system. Hence, materials with a small coherence length will be more affected by quantum fluctuations.

In conclusion, we have measured that in the $Cr_{1-x}Re_x$ system is by the QCP up to temperatures ~600K. In terms of scaled temperatures, studies in heavy-fermion quantum criticality have shown that the quantum critical regime covers an entropy of about 50% of Rln2. On the other hand, the predictions for complex cuprates may also scan high temperatures[8]. However, the fact that quantum effects take place already at such high actual, not scaled, temperatures in elemental, and non complex, chromium is remarkable and unexpected. Finally, we have found that the QCP can determine the state of the system for materials with small intrinsic coherence lengths far away from it, not only in the δ coordinate, but also in temperature T.

**Acknowledgements**

D.C.F. gratefully acknowledges support from the Brazilian agencies CAPES and CNPQ. M.A.C. and M.N-R. acknowledge support from the program Capes-Cofecub Ph664/10. MN-R acknowledges A. Cano for discussions. S. Pairis, B. Fernandez, P. Plaindoux and Th. Fournier have provided essential technical help.

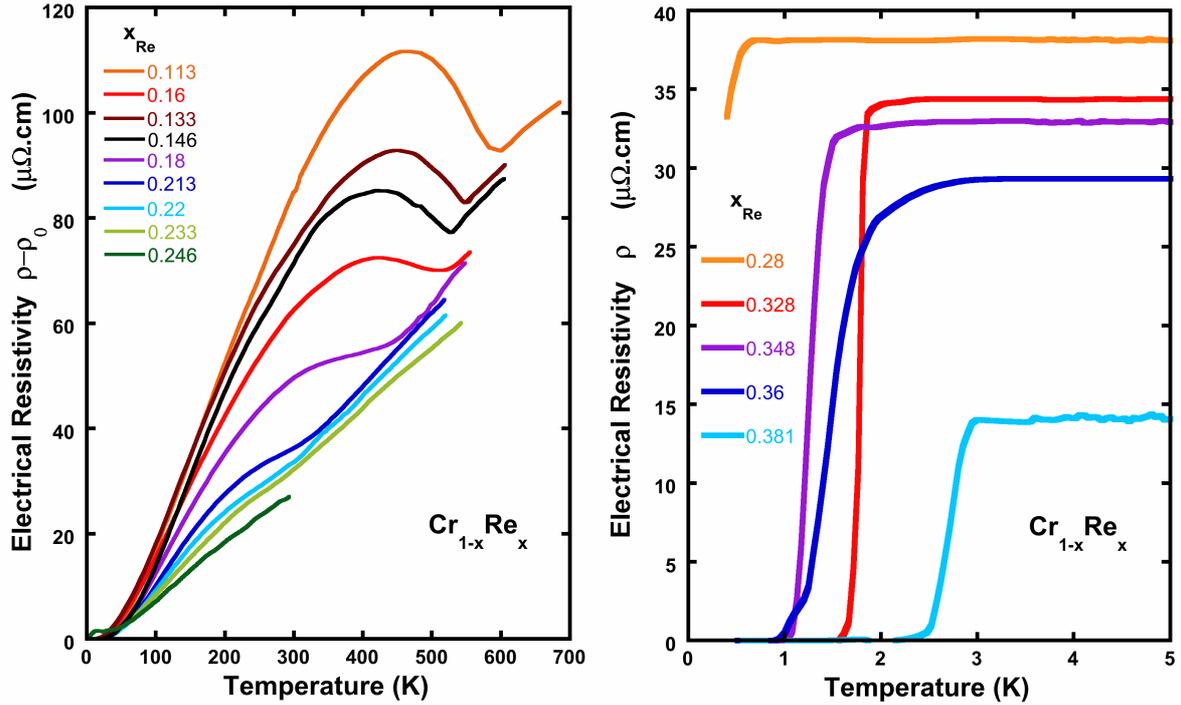

**Figure 1** Electrical resistance for samples with different Re content. Left panel: high temperature behaviour of the low concentration region (with residual resistivity subtracted). The Néel temperature is defined by the steepest increment of the curves, as obtained from their derivative. Right panel: Low temperature, high concentration region, showing the superconducting transitions. The superconducting transition temperature is taken at the onset of the transition.

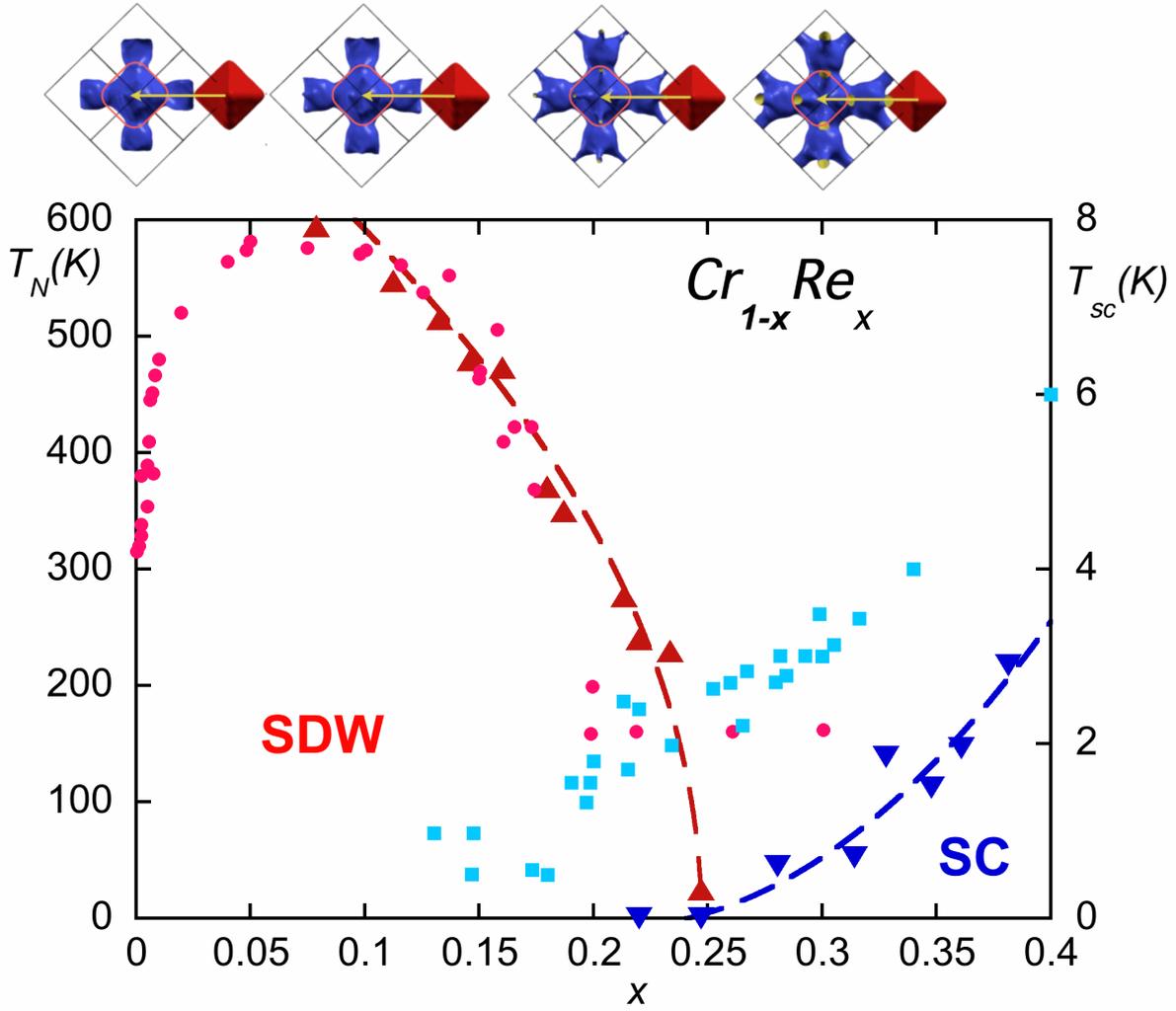

**Figure 2** Lower panel: Phase diagram of $Cr_{1-x}Re_x$ as a function of $Re$ content. Previous determinations of $T_N$ are from Ref. 15 (mauve circles) compared to our data (red triangles). The previously reported superconducting transition temperatures [12] are shown (light blue squares) together with our data (blue inverted triangles). It is clear from the data of our more homogenous samples that superconductivity does not coexist with antiferromagnetism, precluding a non-conventional superconductor. We also observe that the decrease of $T_N$ follows the power law $718[(0.248-x)/0.248]^{0.5}$ from ~600K down to the quantum critical point. Upper Panel: For illustration purposes we show the Fermi surfaces of Cr-Re alloy at x=0, 0.1, 0.2 and 0.3 showing the nesting wavevector that changes form incommensurate (0)

to commensurate (0.1;0.2) and then back to incommensurate (0.3) (details of the calculation in Supp. Mat.).

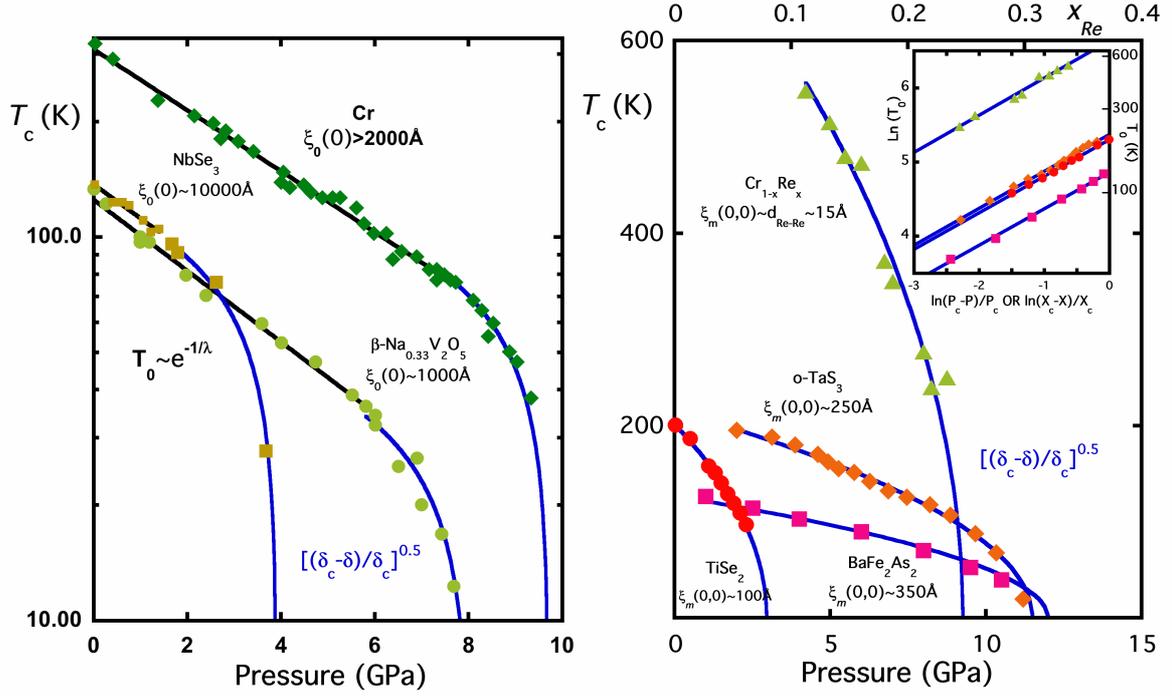

**Figure 3** Left panel: Evolution of the ordering transition temperatures for different materials showing an exponential decrease with pressure of the ordering temperature at low pressures (black lines), followed at high pressures by the $\left[(P_c - P)/P_c\right]^{0.5}$ law (blue lines). $\beta - Na_{0.33}V_2O_5$ (light green circles) [21]; $NbSe_3$ (light brown squares) [19]; $Cr$ (green diamonds) [9]. Right panel: Materials that show the $\left[(\delta_c - \delta)/\delta_c\right]^{0.5}$ ($\delta$ is either pressure or concentration) law in all the measured range. Insert: Log-Log plot to obtain the power of the $\left[(\delta_c - \delta)/\delta_c\right]^\psi$, the slopes of the blue lines give a value of $\psi$ of 0.50, 0.49, 0.49 and 0.50 for $o-TaS_3$ (orange diamonds) [23], $1T-TiSe_2$ (red circles) [22], $BaFe_2As_2$ (mauve squares) [24] and $Cr_{1-x}Re_x$ (light green triangles), respectively. This type of fit yields a very reliable value of the exponent. The values shown for the measured correlation lengths are described in Supplemental material.

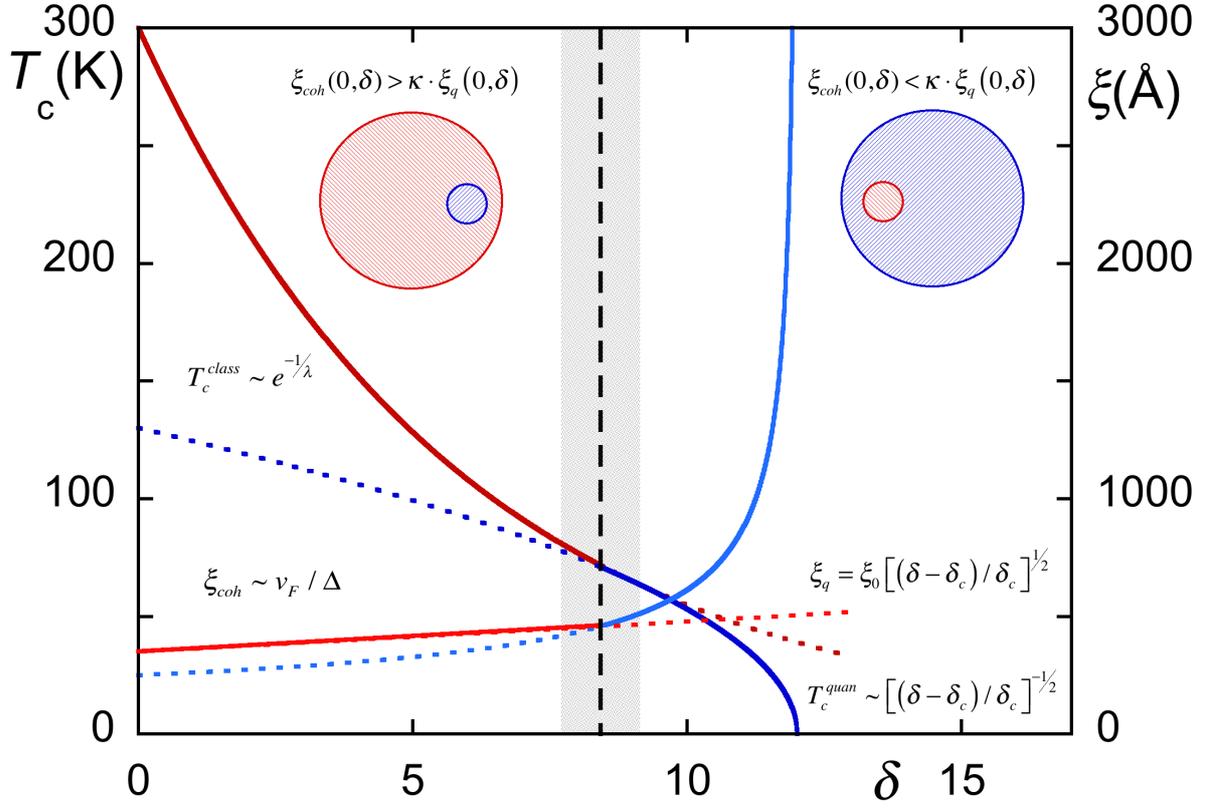

**Figure 4** Schematic phase diagram showing the behavior of a typical ordering transition as a function of an external parameter $\delta$. The crossover from BCS exponential to the quantum power law regime is governed by a crossover ratio $\kappa$ (unity in the figure) of the zero temperature intrinsic correlation length $\xi_{coh}(0,\delta)$ to the quantum correlation length $\xi_q(0,\delta)$.

Table 1

| Compounds | Coherence Length $\xi_m(0,0)$ Å | Method | Ref. |
| --- | --- | --- | --- |
| $Cr$ | 2000 | X-ray diffraction | 28 |
| $NbSe_3$ | 10000 | X-ray diffraction | 25 |
| $\beta - Na_{0.33}V_2O_5$ | 1000 | X-ray diffraction | 27 |
| $o - TaS_3$ | 500 | X-ray diffraction | 29 |
| $1T - TiSe_2$ | 100 | STM | 30 |
| $BaFe_2As_2$ | 350 | Neutron diffraction | 31 |
| $Cr_{1-x}Re_x$ | 15 | Estimated impurity | |